# Beneficial and Harmful Agile Practices for Product Quality

Sven Theobald, Philipp Diebold

Fraunhofer Institute for Experimental Software Engineering,
Fraunhofer-Platz 1, 67663 Kaiserslautern, Germany
{firstname.lastname}@iese.fraunhofer.de

**Abstract.** There is the widespread belief that Agile neglects the product quality. This lack of understanding how Agile processes assure the quality of the product prevents especially companies from regulated domains from an adoption of Agile. This work aims to identify which Agile Practices contribute towards product quality. Hence, data from a survey study is analyzed to identify Agile Practices which are beneficial or harmful for the quality of the product. From 49 practices that were used in the survey so far, 36 were perceived to have a positive impact on product quality, while four practices were rated as being harmful. The results enrich understanding of how product quality can be achieved in Agile, and support selection of practices to improve quality.

## 1   Introduction and Motivation

Agile already is a well-established software development approach, at least in non-regulated domains [1]. Its benefits such as more flexibility and a faster time to market are known. That is why companies from regulated domains such as the automotive or aerospace domains also want to benefit from these advantages. One adoption barrier is the fear of loosing compliance with regulations or certifications, caused by the widespread belief that Agile neglects the quality of the product. Boehm et al. [2] used a dichotomy between agility and discipline, which shows that Agile is not always seen as a disciplined approach. The Agile Manifesto demands: "Individuals and interactions over processes and tools" [3]. This can be misinterpreted as having no defined process. Hence, the quality of the product is often expected to be unpredictable.

This perception is one of many reasons why such companies stick to their traditional approaches with defined and rigorous verification and validation phases. While more flexibility and a faster reaction time to changes are one of the main drivers for Agile adoption [1], dealing with emerging requirements and architectures makes it difficult to plan quality assurance activities upfront and to achieve certifications.

On the other side, Agile processes are reported to produce higher quality [1,4], e.g., based on focusing on a restricted number of most important requirements which are implemented with highest quality. Both different views could be explained by the insight that only the way of assuring quality is different: classical approaches rely on heavy verification and validation activities, while Agile approaches incorporate the realization of quality into the process.

Instead of doing some specific practices (formal reviews, acceptance at certain stage gates, etc.), the contribution towards quality is spread over several practices throughout the whole development process. Therefore it is more difficult to identify which practices are contributing towards quality and to evaluate if a set of practices is sufficient to provide the same trust in the quality as is achieved in traditional approaches. For this reason, it is necessary to understand the contribution of Agile Practices on quality.

Commonly experienced in practice is the phenomenon that people adopt and adapt the method Scrum as their new development process during their Agile transition [1,5]. Scrum is only supposed to be the minimal frame which has to be filled with further development practices. It therefore lacks a description of technical practices which are necessary for a disciplined software engineering. Knowing the effects of those single practices allows to evaluate and improve such a development method. In many cases, a combination of methods and practices is needed, e.g., the use of Scrum [6] enhanced by technical practices from XP [7].

The aim of this work is therefore to identify which elements of Agile, namely which Agile Practices, affect product quality. We use the preliminary results from a survey to analyze which of the practices were perceived by the survey participants as having an impact on product quality.

In Chapter 2, we shortly present the survey which is used as data source, as well as the research questions and analysis approach of this work. The results are presented and discussed in Chapter 3, and we finally conclude the paper together with some suggestions for future work (Chapter 4).

## 2    Research method

In this work, data from a survey study was analyzed. We first present the background information about this survey study to clarify the origin of the data. Afterwards, the research goal is defined and research questions are derived. Finally, the data analysis approach is discussed.

### 2.1    Background

The data used for analysis in this work originate from an ongoing survey study [8]. In this study, the experiences of study participants concerning the impacts of Agile Practices on certain process improvement goals are collected. This is done using A0 posters with a printed matrix of Agile Practices and improvement goals. Participants can place sticky dots in the fields of the matrix, describing that there is an impact of the practice (row) on the improvement goal (column). This impact is rated with the help of color-coding on a scale from strongly positive (green) to strongly negative (red). More information about the research method is provided in [8].

In mid of July, the database contained 1846 data points collected at 17 venues with both academia and industry participants (see **Table 1**). The aggregated results are available on our website [9]. The subset of this data included and discussed in this paper will be those Agile Practices with a (positive or negative) impact on product quality, which is only one of several improvement goals considered in this study.

Table 1. Events, number of collected impacts and participants' information.

| **Events** (ordered by date) | **Impacts** (on Quality/ Overall) | | **Participants** (Background/ No. of) | |
|---|---|---|---|---|
| Agile in Automotive 2016 | 15 | 118 | Practitioners | 170 |
| Profes 2016 | 0 | 143 | Mixed | 150 |
| OOP 2017 | 24 | 76 | Practitioners | 1500 |
| Lean IT Management 2017 | 7 | 78 | Practitioners | 100 |
| AgileXChange 1-2017 | 31 | 112 | Practitioners | 80 |
| AgileLab Copenhagen 2017 | 19 | 123 | Practitioners | 50 |
| Q-Rapids Meeting 2017 | 23 | 83 | Mixed | 20 |
| Agile in Automotive USA 2017 | 20 | 99 | Practitioners | 90 |
| CESI 2017 (@ICSE 2017) | 4 | 22 | Practitioners | 10 |
| XP 2017 | 99 | 513 | Mixed | 280 |
| ScrumDay 2017 | 9 | 68 | Practitioners | 250 |
| EASE 2017 | 9 | 101 | Academics | 90 |
| AgileXChange 2-2017 | 13 | 79 | Practitioners | 80 |
| SPA 2017 | 4 | 26 | Practitioners | 50 |
| Agile Austria 2017 | 33 | 156 | Practitioners | 250 |
| Agile on the beach 2017 | 6 | 37 | Practitioners | 400 |
| ICSSP 2017 | 1 | 12 | Academics | 35 |

### 2.2 Research goal

To overcome the lack of understanding how Agile assures quality, this study aims to identify Agile Practices that have an effect on product quality. The goal of this study was formulated using the GQM template [10]:

> *Identify* Agile Practices *with respect to* their effect on product quality *in the context of* an analysis of preliminary data from a survey study *from the perspective of* Agile practitioners and researchers.

Based on this goal, three research questions (RQs) are defined:
**RQ1:** Which Agile Practices have a positive effect on product quality?
**RQ2:** Which Agile Practices have a negative effect on product quality?
**RQ3:** Which Agile Practices have been rated without a common agreement?

These three research questions help to identify the beneficial and harmful practices, as well as those practices where the impact is varying depending on the context or the implementation of the practice.

### 2.3 Analysis

The existing data (1846 impacts) [9] was filtered to only include the 317 impacts on product quality, which is the improvement goal in the focus of this analysis. The impacts were collected at 16 different events. The reason why one event was missing is that for the first two events, product quality was not included on the poster. At one of those events, the Agile in Automotive (2016), a participant added product quality to the poster in order to report his experiences and we integrated product quality as a standard answer on future posters afterwards. To facilitate analysis, the scale (strongly positive, positive, negative, strongly negative) was transformed into a scale of +2, +1, -1, -2. This enables analysis based on descriptive measures such as the average value.

## 3  Results and discussion

In this chapter, the results of the analysis are presented and discussed along the research questions, followed by a general discussion of results and threats to validity.

From the list of all 49 Agile Practices included in this study, eight practices were not set in relation with product quality at all. The other 41 practices were rated as having an impact on product quality: There were four practices with a negative impact (average <= 0.5), and 36 with a positive impact (average >= 0.5), while one practice, Backlog, received mixed ratings and ended up with a neutral average.

### 3.1  RQ 1 – Beneficial Practices

Overall, 36 of the 49 Agile Practices were perceived to have a positive impact on product quality. Since some of them only received a low number of ratings so far, the trust in the average value is not given for certain practices. Therefore, we provide only those Agile Practices with a more reliable average value in **Table 2**.

Table 2. Most beneficial Agile Practices with at least 10 ratings

| Beneficial Agile Practice | Average | Count | +2 | +1 | -1 | -2 |
|---|---|---|---|---|---|---|
| Test Driven Development | 1.93 | 15 | 14 | 1 | | |
| Pair Programming | 1.83 | 12 | 10 | 2 | | |
| Continuous Integration | 1.77 | 13 | 10 | 3 | | |
| Cross-Functional Team | 1.75 | 28 | 21 | 7 | | |
| Definition of Done | 1.75 | 24 | 18 | 6 | | |
| Definition of Ready | 1.75 | 12 | 9 | 3 | | |
| Unit Testing | 1.64 | 14 | 9 | 5 | | |
| Refactoring | 1.58 | 19 | 14 | 4 | | 1 |
| Backlog grooming | 1.57 | 14 | 10 | 3 | 1 | |
| Iteration Reviews | 1.47 | 19 | 9 | 10 | | |
| Product Owner | 1.4 | 15 | 8 | 6 | 1 | |
| Retrospective | 1.36 | 14 | 9 | 5 | | |
| Scrum Master | 1.23 | 13 | 7 | 4 | 2 | |

For this selection, only practices which received at least 10 ratings were considered in order to increase the reliability of the data. All practices from this list origin from Scrum or XP, most likely because these are the Agile Methods which are most frequently used and known. Sorted by the average rating starting with the practice with the highest average, the list of all other beneficial practices is provided in the following, including the average rating and the number of ratings the practice received:

> Collective Ownership (2/6), Continuous Delivery (2/3), Product Canvas (2/1), Product Vision board (2/2), Work-in-Progress Limit (2/2), Iterative development (1.75/4), Peer Reviews (1.67/3), Self-organizing team (1.67/3), Story Mapping (1.67/3), Personas (1.6/5), Architecture Sprint (1.5/2), Automated Builds (1.5/6), Minimum Viable Product (1.5/6), Shippable Increment (1.5/8), User stories (1.43/7), Coding Styleguides (1.4/5), Sign Up (1.4/5), Communities of Practice (1.33/6), Daily Meeting (1.33/6), Prototyping (1.25/4), Sprint Zero (1/1), Team-Based Estimation (1/1), Relative Estimation (0.5/2).
> 
> *Name (Average/Count)*

### 3.2 RQ 2 – Harmful Practices

Four practices ended up with a negative average, most of them based on a limited amount of ratings (see **Table 3**). It can be seen that velocity is perceived to have a negative contribution towards product quality. Participants rated this impact four times as strongly negative, and three times as negative. In addition, burn charts received a strongly negative rating (three times). The strongly negative ratings for those two practices came all from the XP 2017 conference.

A possible reason for the negative rating of velocity and burn charts could be a misuse of the increased transparency of those practices: If the burn chart shows that the end of the iteration comes closer, or when the current velocity is not as high as in previous iterations, quality is neglected in order to be able to show a better performance. This happens especially when Management uses burn charts or velocity to track the efficiency of the team or even individuals, sometimes also to decide on incentives. This aspect was also mentioned by one of the study participants, when we asked him why he reports a negative impact.

For the other two practices, we cannot tell whether they really have a negative impact or whether these are opinions by individuals. For further analysis, we have to wait for more data.

**Table 3.** Practices with negative impact on product quality

| Harmful Agile Practice | Average | Count | +2 | +1 | -1 | -2 |
|---|---|---|---|---|---|---|
| Taskboard | -1 | 1 | | | 1 | |
| Velocity | -1.57 | 7 | | | 3 | 4 |
| Burn Chart | -2 | 3 | | | | 3 |
| Release planning | -2 | 1 | | | | 1 |

### 3.3 RQ 3 - Ambiguous Practices

While most practices were rated with a clear trend, there are some practices whose ratings are ambiguous. A clear trend means that the practice received either only positive values or only negative ones. In **Fig. 1**, we list all practices that received ambiguous ratings (some positive and some negative ratings).

For all practices with a clear trend, we can assume that, given enough data points, this practice is always beneficial or always harmful for a certain improvement goal, in our case product quality. We expect the participants of our survey to have various context, so we assume that these practices are applicable in different contexts with similar results. Additionally, practices are often adapted to be useable in a certain context or because of lack of knowledge, e.g., most Scrum practices are used with adaptations [5]. It seems, given enough data points, that a practice with a clear trend is always showing certain benefits or drawbacks, independent of the individual implementation and context.

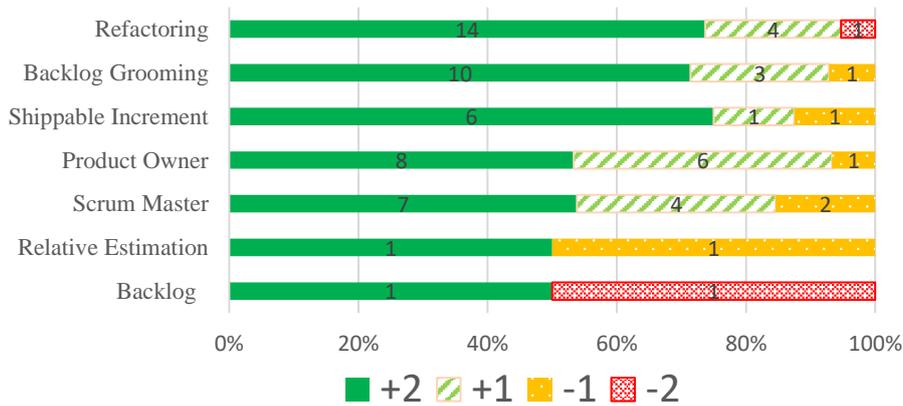

**Fig. 1.** Practices with ambiguous impact on product quality: Distribution of the amount of ratings over the scale (+2,+1,-1,-2).

But if experiences or perceptions of the practices' impacts vary, these practices could be dangerous to adapt or use in certain contexts. Therefore, it is necessary to be aware of those practices during introduction. It is necessary to identify whether those ambiguous ratings are outliers or whether the impact of the practice is really unsteady and context-dependent.

From the data in **Fig. 1**, the negative ratings of the first four practices (Refactoring, Backlog Grooming, Shippable Increment, Product Owner) can be considered as being individual outliers. For the last two practices (Relative Estimation and Backlog), not enough data is available to make any assumption. The Scrum Master received two negative ratings, which tells us that more than one person perceived this Agile role to have a negative contribution towards product quality. Without knowing the reason why those participants reported this harmful impact, we cannot tell whether the role was implemented wrongly, or whether there are certain aspects to this role which really affect quality negatively.

**3.4 Discussion**

The fact that most practices (36 out of 49) were reported as beneficial shows that quality is an important aspect in Agile development. Beside the practices that assure quality directly, such as Unit Testing, Test-Driven Development, Pair Programming and (Code) Reviews, there were other practices contributing towards a higher quality in different ways:

The responsibility for the quality of the product is shared with the help of Agile Practices, such as Collective Ownership, Sign Up/Pull principle, or Cross-functional Team. There is a shared understanding of the quality demands, supported by practices such as Coding Styleguides, Definition of Ready and Definition of Done. Another important aspect how Agile development achieves better quality is to prioritize requirements and focus to only build what the user really needs with a higher quality, using Minimum Viable Product, Product Owner, User Stories, and Personas. To do this, fast feedback cycles (e.g., with Iteration Reviews, Shippable Increment) and continuous improvement (e.g. with Retrospectives) are needed.

All those different ways of incorporating quality into the process indicate that Agile strives for a culture of focusing on the product quality. Discipline is demanded, also to select the right development practices. Only implementing Scrum as the development process might neglect quality, since beneficial practices from other methods like XP are missing.

Not many practices were rated negatively. The only alarming practices were the use of burn charts and velocity, which are frequently used together with Scrum. Therefore, many practitioners are affected and should check whether their implementations of these practices have a negative impact on quality in their specific context.

The threats identified in [8] discuss the validity of the data. The main threat is converting the ordinal scale to an interval scale for easier comparison using the average value. Therefore the average values do not have a high explanatory power. Instead of considering the detailed ranking, only the tendency towards positive or negative impact should be considered. The proposed practices are by no means complete, since not all available Agile Practices were covered by the survey. Only a limited amount of ratings were given so far, so the validity of results needs to be improved by increasing the size of the data set, which is continuously done with the ongoing survey.

## 4 Conclusion and Future Work

An analysis of the preliminary results [8] a survey study [8] showed that there exist many Agile Practices which contribute towards product quality. Out of the 49 practices used in the survey, 36 practices were reported as beneficial (RQ1). On the other hand, only four practices were reported with only a few ratings as being harmful for product quality (RQ2). Practitioners should especially be careful with the practice Velocity. There is very little disagreement on which practices contribute positively or negatively (RQ3). Thus, both academics and practitioners seem to have a common perception of the impacts of Agile Practices. Despite the assumed individual context variations and adaptations of practices, most practices show a stable impact.

The identified practices help to better understand how Agile addresses product quality. This knowledge can be used to select dedicated Agile Practices for adoption. Knowing which practices contribute to product quality facilitates a mapping of Agile Practices and certain regulations in order to identify how Agile approaches fulfill the requirements of such standards concerning the achievement of high quality products.

Since this is a continuous survey study, we rely on support by event organizers or event visitors who want to place our poster to collect experiences from the participants. This is your chance to contribute to the Agile community to increase our understanding of the impacts of Agile Practices. With a growing number of reported impacts, the validity of our analysis can be increased.

Additionally, qualitative statements have to be collected to provide some potential reasons for the reported impacts, e.g. in case studies or interviews. Further, the high-level improvement goals can be further refined to provide insights. In the case of product quality, future work could investigate which products or product parts and which refined quality aspects are addressed.

**Acknowledgements**. This work was partly funded by the German Federal Ministry of Education and Research in a Software Campus project (BMBF 01IS12053) and as part of the research project ProKoB[1] (BMBF 01IS15038).

---

[1] www.prokob.info